\documentclass[preprint]{aastex}

\newcommand\simlt{\lower.5ex\hbox{$\; \buildrel < \over \sim \;$}}

\begin{document}
\title{Calorimetry of gamma-ray bursts: echos in gravitational waves}
\author{Maurice H.P.M. van Putten}
\affil{Department of Mathematics,
Massachusetts Institute of Technology, Cambridge, MA 02139-4307}
\author{Amir Levinson}
\affil{School of Physics and Astronomy, Tel Aviv University, Tel Aviv, Israel}

\begin{abstract}
  Black holes surrounded by a disk or torus may drive 
  the enigmatic cosmological gamma-ray bursts (GRBs). 
  Equivalence in poloidal topology to pulsar magnetospheres shows
  a high incidence of the black hole-luminosity $L_H$ into the surrounding
  magnetized matter. We argue that this emission is re-radiated into 
  gravitational waves at $L_{GW}\simeq L_H/3$ in frequencies of order 1kHz, 
  winds and, potentially, MeV neutrinos.
  The total energy budget and input to the GRB from baryon poor jets 
  are expected to be standard in this scenario, consistent with recent 
  analysis of afterglow data. 
  Collimation of these outflows by baryon rich disk or torus winds 
  may account for the observed spread in opening angles up to about
  $35^o$. This model may be tested by future LIGO/VIRGO observations.
\end{abstract}


\section{Introduction}

It is now widely believed that progenitors of GRBs are 
massive stars or black hole-neutron star binaries. These systems
have ample and well-defined energy in angular momentum to produce 
the relativistic outflows which drive GRBs. Support for hypernovae 
in massive stars \citep{woo93,pac97,pac98,bro00} derives from an association 
with star-forming regions \citep{pac98,blo00}, the GRB 980425/SN 1998bw event 
(although this source exhibits markedly different properties than a typical 
burst) and a potential GRB/Soft X-ray transient connection \citep{bro00,bro01}, 
e.g., as in GRO J1655-40 \citep{isr99} and V4641 Sgr \citep{oro01}. Chemical 
abundances in the latter have been attributed to the intercept of hypernova-debris 
by the secondary during the GRB-event. A likely outcome of the collapse of a 
massive star is a black hole surrounded by a disk or torus.
The coalescence 
of black hole-neutron star binaries may likewise produce black hole-torus 
systems when the black hole spins rapidly \citep{pac91}. 
This remains of interest 
in view of their occurence rate \citep{phi91,nar91,bro99} which, 
although marginally, 
is consistent with GRB statistics set by estimates of the 
beaming factors (see below).

In this {\em Letter}, we classify the major channels of radiation from 
black hole-torus systems and provide estimates for their net fluences
(see \cite{mvp01} for a review).
Outflows powered directly by the black hole
enables baryon poor input to the GRBs.
Radiation from the torus establishes channels for ``unseen" emissions:
gravitational radiation, Poynting flux-dominated
and baryonic collimating winds and, possibly, neutrino emissions. In the
GRB/SXT-association of \cite{bro00,bro01}, these emissions are augmented
by hypernova disk-winds which provide a
chemical deposition of metals onto the companion star.
We shall find that a dominant fraction of the energy will be liberated
in gravitational waves.
As predictions for future calorimetry in multi-window observations,
estimates of the fluences in these ``unseen" emissions 
may provide the most stringent observational test
of the black hole-torus system. 
As an observational test for the black hole-torus
association to gamma-ray bursts, we may further consider recovering the
the present bi-modal distribution in
durations in the BATSE catalogue in these unseen radiation channels.

Recent analysis of afterglow data in GRBs with measured redshifts appears to 
indicate a standard energy release of prompt gamma rays in long bursts, and a 
rather wide range of observed beaming factors \citep{fra01}.  
Below, we argue that standard energy release 
and a range of collimation factors 
is a natural consequence of black hole-torus systems 
in a suspended accretion state, which we associate with long bursts
\cite{mvp01b}.  

In \S 2, the basic features of black hole plus disk or torus systems
is summarized. \S 3 discusses the clustering-and-spread in leptonic
outflows from black hole-torus systems. \S4 presents an outlook for
unseen emissions in gravitational waves.

\section{Non-thermal emissions from black hole-torus systems}

The black hole-torus system represents hypernovae, or the expected outcome of
black hole-neutron star coalescence. The torus forms as remnant matter of
a fallback envelope of a massive star or as the debris of a neutron star.
A net magnetic flux carried by the torus supports a similarly shaped
magnetosphere. Surrounded by a magnetic field with net poloidal magnetic
flux, the central black hole assumes an equilibrium
magnetic moment \citep{mvp01}
\begin{eqnarray}
\mu_H^e\sim aBJ_H,
\label{EQN_MU}
\end{eqnarray}
where $a=J_H/M$ denotes the specific angular momentum of a black hole with 
angular momentum $J_H$ and mass $M$, and $B$ the field-strength of the 
torus magnetosphere. The magnetic moment given by eq. (\ref{EQN_MU}) is 
no ``fourth hair" of the black hole \citep{car68}: it is generated by an 
equilibrium charge $q^e\sim BJ_H$ of the black hole \citep{wal74,dok87,lee01}.
This magnetic moment serves to regulate an essentially uniform and maximal 
horizon flux, preserved at arbitrary rotation rates. This permits the horizon
to couple to the inner face of the torus and to support of an open flux-tube
to infinity \citep{mvp01}. The net horizon luminosity $L_H$ is, correspondingly,
the sum of a luminosity $L_T$ into the torus and a luminosity $L_p$ to infinity.

Equivalence in poloidal topology to pulsar magnetospheres indicates
a high incidence of the black hole-luminosity -- 
by way of Maxwell stresses -- into the 
surrounding magnetized matter
(adapted from \cite{gol69,tho86,mvp99a,bro01})
\begin{eqnarray}
L_{T}=\omega_T(\omega_H-\omega_T)f_T^2A_H^2,
\label{EQN_LT}
\end{eqnarray}
where $f_T$ denotes the fraction of poloidal horizon flux $2\pi A_H$ 
which connects to the torus, and $\omega_T,\omega_H$ denote the angular 
velocities of the torus and the black hole, respectively. A
suspended accretion state arises when $\omega_H/\omega_T$ is 
sufficiently large \cite{mvp01b}.

Leptonic outflows are produced by rotating black holes in
an open flux-tube along their axis of rotation, in response to 
differential frame-dragging 
\citep{mvp00a,hey01,mvp01}. 
This open flux-tube is endowed with conjugate radiative-radiative boundary 
conditions, at the horizon and finity.
Global closure of the current -- asymptotically null
in the force-free sections on the horizon and infinity --
over an outer flux-tube supported by the torus
gives rise to a net luminosity in $e^\pm\gamma$
given by \citep{mvp01}
\begin{eqnarray}
L_{p}=\frac{1}{2}\omega_T(\omega_H-2\omega_T)f_o^2A_H^2,
\end{eqnarray}
where $f_o$ denotes the fraction of poloidal horizon flux which supports the 
open flux-tube. 
This in situ creation of leptonic outflows takes place whenever the
black hole rotates faster than {\em twice} the angular velocity of a torus
in prorate rotation.

  The energetic coupling (\ref{EQN_LT}) permits a high luminosity
  $L_{GW}$ in
  low-frequency gravitational-wave emissions from the torus, by
  generic non-axisymmetric deformations. A time-averaged luminosity can be 
  estimated in the suspended accretion
  state \cite{mvp01b}.
  When most of the magnetic field on the horizon is anchored to the surrounding
  matter, i.e. $f_T\simeq1$,
  we have $L_H\simeq L_T$ and hence
  \citep{mvp01}
  \begin{eqnarray}
  L_{GW}\simeq \omega_T^2A_T^2\simeq L_H/3,
  \label{EQN_GW1}
  \end{eqnarray}
  where $2\pi A_T$ denotes the net poloidal flux in the torus. 
  Here, $\omega_T/\omega_H\simeq f_H^2/3$ in terms of the fraction
  $f_H$ of the flux supported by the torus which connects to the 
  horizon of the black hole
  \citep{mvp00a}.

  An appreciable fraction of (\ref{EQN_LT}) may dissipate in the disk, 
  probably giving rise to emission of neutrinos if maintained 
  at high enough temperatures, and perhaps other forms of radiation.
  The torus may loose baryon rich material at a 
  substantial rate. This may provide a baryonic collimating
  wind launched 
  along the open flux tubes (see below), and a chemical deposition
  of metals onto the companion star. 

To summarize, the work performed by the black hole as it sheds off its angular momentum
in the course of the evolution to a lower energy state (cf. the Rayleigh criterion),
is liberated through two important channels \citep{mvp01}:

\newcounter{bean}
\begin{list}
{{\Roman{bean}}}{\usecounter{bean}
 \setlength{\rightmargin}{\leftmargin}}
\item  
  Baryon poor outflows along open flux-tubes supported by the spin-energy
  of the black hole \citep{mvp00a,hey01,mvp01} that, we conjecture, provide
  the free energy source that powers GRBs. This work is proportional to
  $f_o^2=(\Omega_H/4\pi)^2<<1$, whereby the remaining fraction $f_T\simeq1$.
  The important GRB emissions form a small 
  fraction of the total luminosity $L_H$ of the black hole.
\item  
  Gravitational wave-emissions from non-axisymmetric
  deformations in the torus, powered by about one-third
  of the black hole luminosity.
  These frequencies are 
  correlated with the Keplerian
  angular velocity of the torus and, hence, of the order of 1kHz
  --  low frequencies relative to the quasi-normal
  mode oscillations of the black hole.
\end{list}

In the hypernova model of \cite{bro00,bro01}, a third
channel consists of disk winds providing chemical depositions 
of metals onto the companion star. The energy $\sim 10^{52}$ erg
in these winds \citep{mae01} may derive from $L_T$ by a safe
margin (see below). Unknown is the
role of analogous baryonic winds 
from black hole-torus systems formed in mergers of
black hole-neutron star binaries.

\section{The true energy budget and clustering-and-spread in outflows}

The fraction of the hole rotational energy liberated as leptonic outflows is
approximately $(\Omega_H/4\pi)^2 << 1$, where $\Omega_H$ is the solid angle on 
the horizon occupied by an open flux-tube \citep{mvp01}.  
The remaining spin down 
energy is deposited in the torus as described in (II) above.  
By the first-law of black hole-thermodynamics, the
efficiency of delivering spin-energy into the surrounding torus is 
given by the ratio
of angular velocities $\omega_T/\omega_H=2/((R/M_{H})^{3/2}+1)$ of the 
torus and black hole,
respectively, for a maximally spinning black hole. Here, $R$ denotes the
radius of the torus and $M_{H}$ the mass of the black hole, assuming 
the torus to be in approximately Keplerian motion.  For a
conservative
value $\omega_T/\omega_H\sim 0.1$, the available black hole energy of a 
maximaly rotating black hole is
\begin{eqnarray}
E_{H}\simeq 4\times 10^{53}(M_{H}/7M_\odot)\ \ {\rm ergs}.
\label{E_BH}
\end{eqnarray}
Here, the fiducial scale of $7M_\odot$ is motivated by the range of about
$4-14M_\odot$ in dynamically determined masses of black hole candidates in
X-ray novae (see \cite{mvp01} for a recent update).

The suspended accretion state is expected to result in a relatively thick torus,
as it will be hot and in a state of super-/sub-Keplerian motion at the inner/outer face 
due to the powerful competing torques acting on it. This promotes a radially slender and a
poloidally extended shape. It should be mentioned that the torus mass and  
magnetic-to-kinetic energy ratio are the two main uncertain parameters, 
also in regards to the time-scale of the suspended accretion state \citep{mvp01b}.

A relatively thick torus produces a funnel surrounding the black hole.
The horizon solid angle $\Omega_H$ of the open flux-tube on the horizon will 
depend on the poloidal structure of the torus magnetosphere surrounding the black hole 
-- as supported by the inner face of the torus in its immediate vicinity.  Tori with 
inner faces larger than the black hole size are expected to give rise to a similar poloidal 
topology of the torus magnetosphere surrounding the black hole, and hence similar 
solid angles $\Omega_H$.  This leads to the expectation that both, the total black
hole energy, $E_H$, and the fraction emerges along the open flux tubes, $E_j$, 
should be standard. 
Adopting the estimate by \citep{fra01} for the true energy of prompt 
emission (in bi-polar outflows): $\bar{E}_{\gamma} \simeq 5\times10^{50}$ ergs, we obtain 
\begin{eqnarray}
\Omega_H/4\pi=(E_j/E_H)^{1/2}\simeq 0.1 (M_H/7M_\odot)^{-1/2}(\epsilon/0.1)^{-1/2},
\label{Omga_H}
\end{eqnarray}
where $\epsilon$ denotes the conversion efficiency of bulk energy of the leptonic outflow 
into gamma rays.  While existing models vary in their estimates of $\epsilon$, 
an approximate mean value of about $0.15$ appears to be reasonable for GRB 
emission from internal shocks \citep{kob97,dai98,gue00,pai00}. Radiative viscosity 
mediated through the agency of some background radiation field (a component that 
may be particularly relevant for the class of models considered here) may enhance 
the efficiency of internal shocks considerably \citep{lev98}.  Alternatively, high 
efficiency (as well as a preferable $\nu F_{\nu}$ peak of prompt emission) can be 
naturally achieved in compact fireball models, whereby prompt GRB gamma rays
near the peak of the spectral energy distribution are produced on compact scales,
prior to the acceleration of the fireball to its terminal Lorentz factor \citep{eich00}.
Thus, a value of $\Omega_H/4\pi\sim 0.1$, corresponding to an opening angle
\begin{eqnarray} 
\theta_H\simeq 35^\circ,
\label{EQN_A}
\end{eqnarray} 
seems reasonable.  Given the 
observed mean beaming factor, $\bar{f}_b\sim 2\times10^{-3}$, this generally 
implies that the baryon poor outflow should be further collimated into a solid 
angle $\Omega_j\sim \Omega_H/30$.  Further collimation 
may proceed through interactions with the baryon rich wind surrounding it \citep{lev00}.  
If the disk or torus is initially hot (at MeV temperature) and dense, as anticipated
if it forms as a result of collapse of a massive star or coalescence 
of compact objects, then it will first cool down over a timescale of several seconds 
via neutrino emission.  During this process a 
fraction $\eta$ (a few percent) of a solar mass will be blown off, carrying a total 
energy of order $\eta M_{\sun}c^2$ \citep{lev93}.  Additional mass loss may arise from 
the interaction of the black hole and the torus, as mentioned above.

The degree of collimation depends on the parameters of the baryonic wind, which in 
turn might be sensitive to the conditions in the disk (as well as its structure) and is, 
therefore, expected to exhibit large variations. This can explain the diversity of 
beaming factors exhibited by the sample studied in \citep{fra01}. Preliminary 
analysis \citep{lev00} shows that the opening angle of the fireball thereby collimated 
is proportional to the ratio of luminosities of the baryon poor and baryon rich outflows.
For a GRB having a beaming factor of order the mean found by \citep{fra01};
e.g., GRB 990123, we find that the energy $E_{cw}$ on the confining wind is
roughly 
\begin{eqnarray}
E_{cw}\simeq (2f_b)^{-1/2} E_j \epsilon^{-1} \simeq 5\times 10^{52}\mbox{~ergs},
\end{eqnarray}
below the total energy deposition in 
the torus by a safe margin (see eq. [\ref{E_BH}]).
Further work is needed to quantify the relation between the collimation 
factor and the luminosity of the baryonic wind for a broader range of physical 
conditions, 
and for different assumptions on the boundary conditions. 

\section{Gravitational-wave bursts from black hole-spin in long GRBs}

There are at least five aspects to a black hole-torus system which
suggest considering their potential as LIGO/VIRGO sources of 
gravitational radiation, produced by emissions
from the torus. 
  By equivalence in poloidal topology to pulsar
magnetospheres, the torus is strongly coupled to the spin-energy of 
the black hole \citep{mvp99a,mvp01b};
  non-axisymmetric deformations in the torus will produce 
gravitational and radio emissions (modulations) at low frequencies
(relative to those of the quasi-normal modes of the horizon),
e.g., at twice the Keplerian frequency if produced by
lumpiness;
  compact magnetized objects which reach their
Schwarzschild radius with gravitationally weak magnetic fields,
radiate predominantly in gravitational rather than electromagnetic waves
(cf. young pulsars, e.g., \cite{sha83});
  non-axisymmetric deformations in the torus are expected by instabilities,
such as from self-gavity 
and consistent with the
requirement for intermittency at the source \citep{pir98};
  the true rate of GRBs should be frequent, as inferred from their large 
beaming factor \citep{fra01}.

The gravitational wave-luminosity can be calculated in
a suspended accretion state \citep{mvp01}.
The frequencies will be, to leading order, related
to the Keplerian angular velocity of the torus.
On the secular time-scale of spin-down of the black hole,
these frequencies will trace a horizontal branch 
in the $\dot{f}(f)-$diagram \citep{mvp00d}. In particular,
lumpiness in the torus
produce a gravitational-wave frequency 
at twice the Keplerian frequency \citep{mvp01}:
\begin{eqnarray}
f_{gw}(t)\sim1-2\mbox{kHz}/(1+z),~~~\mbox{d}f_{gw}(t)/\mbox{d}t=\mbox{const.} 
\label{EQN_F}
\end{eqnarray}
Here, canonical values are used for 
a black hole-torus system at redshift $z$. 
Other frequencies may arise as described in studies
of QPOs (cf. \cite{stel00}).
If the torus is 
unstable against breaking up in clumps, or if the torus shows violent 
expansions in its 
mean radius, the gravitational waves will be episodical, and will correlate 
with sub-bursts in long GRBs. It should be mentioned that the horizontal branch may be above 
or below the $f-$axis, corresponding to a positive or negative linear chirp in the frequency 
dynamics. This uncertainty results from the uncertainty in the details of the radial dependence
of the fraction of interconnecting field-lines between the black hole and the inner face of 
the torus as a function of major radius of the torus.

The torus is expected to be luminous also in winds and, when
sufficiently hot, neutrino emissions. The energy liberated in
gravitational radiation, winds, and neutrinos may be roughly equal 
\citep{mvp01}. 
By eq. (\ref{EQN_GW1}),
the black hole luminosity (eq. [\ref{E_BH}]) into the 
torus serves as an estimate for the energy released in 
gravitational waves:
\begin{eqnarray}
E_{gw}\simeq 10^{53} (M_H/7M_\odot)\ \ {\rm ergs}.
\label{EQN_GW}
\end{eqnarray}
Since gravitational-wave emission is essentially unbeamed, the beaming
factor $\sim 500$ found by \cite{fra01} suggests that long GRBs, 
observed at redshifts of order unity, take place as gravitational wave-bursts 
at a rate of few$\times 10^5$/year, or a few times per year 
within a distance of 100Mpc. With their gravitational 
emissions (eq. [\ref{EQN_GW}]) in the frequency range given by eq. 
(\ref{EQN_F}) this suggests 
that long GRBs are potentially powerful LIGO/VIRGO burst-sources of 
gravitational radiation \citep{mvp01}. 

The above suggests at least three aspects to future LIGO/VIRGO observations.
LIGO/VIRGO may quantitatively probe the inner most structure of black hole-torus
systems by tracking the secularly evolving frequencies. 
A link to GRBs may be established by comparing 
duration statistics from these LIGO/VIRGO detections with the BATSE 
catalogue (after redshift corrections). A link to a binary progenitor
system may be made by searching for a progenitor chirp to the emissions
from the torus. If present, it would be of interest to consider 
extracting the individual binary masses by combining the chirp mass
from precursor emissions with the black hole mass determined from the
(secularly evolving) frequencies from the torus.

\section{Summary and Conclusions}

A black hole surrounded by a disk or torus is a likely outcome of the collapse
of a massive star or coalescence of compact objects and may power the
enigmatic GRBs.  
The rotational energy of the black hole should be released 
both along the axis of rotation and into the equatorial plane
owing to its interaction with the surrounding disk or torus.  
Baryon poor outflows are expected to be produced along the axis of
rotation, while Maxwell stresses dominate the energetics onto the
surrounding magnetized matter. A torus will re-radiate the latter
in gravitational waves, winds and, possibly, neutrinos. 

Sufficiently thick tori should yield a similar poloidal structure of the
magnetosphere in the neighborhood 
of the black hole and, hence, roughly similar opening angles of the 
outflow on the horizon, leading to the extraction a standard 
fraction of the spin-energy in the form of outflows that power GRBs.
This can quite naturally account for the 
inferred clustering of GRB energies.  
The initial cooling of the torus, and subsequent 
dissipation of some fraction of the unseen hole energy, 
will drive considerable mass 
loss of baryons from the torus.  
Collisions of the baryon poor outflows and the baryon 
rich winds can provide a collimation mechanism. 
As the degree of collimation depends 
on the wind parameters, and might be sensitive to the specific conditions 
in the torus, 
it should lead to a range of opening angles for the collimated leptonic 
outflows, 
which may explain the observed spread of beaming factors.
The same power input to the disk or torus is expected to
expell baryonic winds in the equatorial plane, which may account
for the chemical depositions in metals onto the companion star
\citep{bro00,bro01}.

A standard total hole energy and the fraction powering the associated 
GRB determines the solid angle 
$\Omega_H$ occupied by the open 
flux tubes near the horizon. 
This relates to gravitational waves as a potentially new observable,
and to the true energy of
the prompt GRB emission.   
If $\Omega_H$ is indeed similar in all sources, 
it will appear as a break in the distribution of opening angles
(cf. (\ref{EQN_A})) that may be observable in 
a large enough sample, and can serve as an important test for this 
model. 

A major fraction of the rotational energy of the black hole is predicted to be
liberated in gravitational waves which overlap with the bandwidth of
advanced LIGO/VIRGO. 
Determining this power output promises
true calorimetry of black 
hole-torus systems.

{\bf Acknowledgement.} This research is supported by NASA Grant 5-7012, 
an MIT C.E. Reed Fund, a NATO Collaborative Linkage Grant, and a grant 
from the Israel Science Foundation. The
authors thank S. Kulkarni and the referee for constructive
comments, and P. Meszaros for drawing attention to QPOs.


\begin{thebibliography}{}
\bibitem[Brown et al.(1999)]{bro99} Brown, G.E.,
  Lee, C.-H., \& Bethe H.A., 1999, NewA, 4, 313
\bibitem[Brown et al.(2000)]{bro00} Brown, G.E., Lee, C.-H., Wijers,
  R.A.M.J., Lee, H.K., Israelian, G., \& Bethe, H.A., 2000, NewA, 5, 191
\bibitem[Brown et al.(2001)]{bro01} Brown, G.E., Bethe, H.A., \& Lee, H.-K.,
  Selected Papers: Formation and Evolution of Black Holes in the Galaxy
  (World Scientific), in preparation (See commentary on Brown et al.(2000)).
\bibitem[Bloom et al.(2000)]{blo00} Bloom J.S., Kulkarni S.,
  and Djorgovski S.G., 2000, AJ, subm; astro-ph/001076.
\bibitem[Carter(1968)]{car68} Carter B., 1968, Phys. Rev., 174, 1559
\bibitem[Dokuchaev(1987)]{dok87} Dokuchaev V.I., 1986, Sov. Phys. JETP, 65, 1079
\bibitem[Eichler \& Levinson(2000)]{eich00} Eichler D. \& Levinson A., 2000, ApJ, 
529, 146
\bibitem[Frail et al.(2001)]{fra01} Frail et al., 2001,
  astro-ph/0102282
\bibitem[Goldreich and Julian(1969)]{gol69}
  Goldreich, P., and Julian, W.H. 1969, \apj, 157, 869
\bibitem[Guetta et al.(2000)]{gue00} Guetta D., Spada M., \& Waxman E., 
  2001, ApJ, submitted; astro-ph/0011170 
\bibitem[Heyl(2001)]{hey01} Heyl J.S, 2001, Phys. Rev. D, 63, 064028
\bibitem[Israelian et al.(1999)]{isr99} Israelian, G., Rebolo, R.,
  Basri, G., Casares, J., \& Mart\'in, E.L., 1999, Nature, 401, 142
\bibitem[Kobayashi et al.(1997)]{kob97} Kobayashi, S., Piran, T., \& Sari,
  R., 1997, ApJ, 490, 92 
\bibitem[Kouveliotou et al.(1993)]{kou93} Kouveliotou, C., et al., 1993, ApJ, 413, L101
\bibitem[Lee et al.(2001)]{lee01} Lee H.K., Lee C.-H., \& van Putten M.H.P.M., MNRAS, to appear
\bibitem[Levinson(1998)]{lev98} Levinson, A., 1998, ApJ, 507, 145
\bibitem[Levinson \& Eichler(1993)]{lev93} Levinson, A., \& Eichler, D.,
  1993, ApJ, 418, 386
\bibitem[Levinson \& Eichler(2000)]{lev00} Levinson, A., \& Eichler, D.,
  2000, Phys. Rev. Lett.,
  85, 236
\bibitem[Daigne \& Mochkovitch(1998)]{dai98} Daigne, F., \& Mochkovitch, R.,
 1998, MNRAS, 296, 275
\bibitem[Maeda et al. (2001)]{mae01} Maeda, K., et al., 2001, ApJ Lett.,
 submitted; astro-ph/0011003
\bibitem[Narayan et al.(1991)]{nar91} Narayan, R., Piran, T., \& Shemi, A., 1991, ApJ,
   379, L17
\bibitem[Orosz et al.(2001)]{oro01} Orosz, J.A., Kuulkers, E., McClintock,
   J.E., Garcia, N.R., Jain, R.K., Bailyn, C.D., \& Remillard, R.A., ApJ, 2001, submitted
\bibitem[Paciesas et al.(1999)]{pac99} Paciesas, W.S., et al., 1999, ApJS, 122, 465
\bibitem[Paczy\'nski(1991)]{pac91} Paczy\'nski, B.P. 1991, Acta Astron., 41, 257
\bibitem[Paczy\'nski(1997)]{pac97} Paczy\'nski, B.P. 1997, astro-ph/9706232
\bibitem[Paczy\'nski(1998)]{pac98} Paczy\'nski, B.P. 1998, \apjl, 494, L45
\bibitem[Panaitescu \& Kumar(2000)]{pai00} Panaitescu, A., \& Kumar, P.,
  2000, ApJ, 543, 66
\bibitem[Papadopoulos \& Font(2001)]{pap01} Papadopoulos, P., \& Font, J.A.,
  2001, Phys. Rev. D., 63, 044016
\bibitem[Phinney(1991)]{phi91} Phinney, E.S., 1991, ApJ, 380, L17
\bibitem[Piran(1998)]{pir98} Piran, T.,
  1998, Physics Reports, 314, 575; ibid., 1999, 333, 529
\bibitem[Shapiro and Teukolsky(1983)]{sha83} Shapiro S.L. and Teukolsky S.A.,
  Black holes, white dwarfs and neutron stars (Wiley-Interscience), p283
\bibitem[Stella (2000)]{stel00} Stella, L., 2000, 
  in Proc. X-ray Astron. 1999 - Stellar
  Endpoints, AGN and the 
  Diffuse Background, G. Malaguti, G. Palumbo \& N. White (eds), 
  (Gordon and Breach, Singapore)
\bibitem[Thorne et al.(1986)]{tho86} Thorne, K.S., Price, R.H. \& McDonald, D.A.,
  1986, Black Holes: The Membrane Paradigm (New Haven: Yale University Press)
\bibitem[van Putten(1999)]{mvp99a} van Putten, M.H.P.M., 1999, {Science}, 284, 115
\bibitem[van Putten(2000a)]{mvp00a} van Putten, M.H.P.M., 2000a, 
  Phys. Rev. Lett., 84, 3752; ibid. submitted
\bibitem[van Putten \& Ostriker(2001)]{mvp01b}
  van Putten, M.H.P.M., and Ostriker, E.C., 2001, ApJ, 522, L31
\bibitem[van Putten and Sarkar(2000)]{mvp00d} van Putten, M.H.P.M., and
  Sarkar A., 2000, Phys. Rev. D, 62, 041502(R)
\bibitem[van Putten(2001)]{mvp01} van Putten, M.H.P.M., 2001, Physics Reports, in press
\bibitem[Wald(1974)]{wal74} Wald, R.M., 1974, Phys. Rev. D., 10, 1680
\bibitem[Woosley(1993)]{woo93} Woosley, S.E. 1993, \apj, 405, 273
\end{thebibliography}
\end{document}